\documentclass{article}

\usepackage{amssymb}
\usepackage{amsmath}
\usepackage{latexsym}
\newtheorem{theorem}{Theorem}
\newtheorem{lemma}[theorem]{Lemma}
\newtheorem{proposition}[theorem]{Proposition}

\newtheorem{fact}[theorem]{Fact}

\newenvironment{proof}[1]{%
  \begin{trivlist}{}{\setlength{\topsep}{0cm}\setlength{\partopsep}{0cm}}
  \item \textbf{#1.\@}\hspace*{1ex}\ignorespaces}%
  {\makebox[0cm]{}\nolinebreak\hfill$\Box$\end{trivlist}}

\def\ket#1{{|{#1}\rangle}}

\def\glb1q{\diamond}

\def\R{{\mathbb R}}
\def\C{{\mathbb C}}

\def\gates{{\Gamma}}
\def\Hom{\mbox{Hom}}
\def\End{\mbox{End}}
\def\Monodr{\mbox{\cal M}}

\def\qudit{{qu$d$it}}
\def\qudits{{qu$d$its}}

\begin{document}

\title{Deciding universality of quantum gates}
\author{
{G\'abor Ivanyos}\thanks{
{Computer and Automation Research Institute
of the Hungarian Academy of Sciences,
Kende u. 13-17, H-1111 Budapest, Hungary.}
{E-mail: \tt Gabor.Ivanyos@sztaki.hu.}
{Research partially supported by
the project RESQ IST-2001-37559 of the IST-FET program of the EC, 
and by the Hungarian Scientific Research Fund (OTKA) under
grants T42706 and T42481. 
}
}
}

\maketitle

\begin{abstract}
We say that collection of $n$-qudit
gates is universal if there exists
$N_0\geq n$ such that for every $N\geq N_0$
every $N$-qudit unitary operation
can be approximated with arbitrary precision
by a circuit built from gates of the
collection. Our main result is an upper bound
on the smallest $N_0$ with the above property. 
The bound is roughly $d^8 n$, where
$d$ is the number of levels of the base system
(the '$d$' in the term \qudit{}.)
The proof is based on a recent result on 
invariants of (finite) linear groups.
\end{abstract}

\section{Introduction}

A \qudit{} is a vector of norm 1
from the Hilbert space $\C^d$,
an $n$-\qudit{} state is an element 
of norm 1 of $(\C^d)^{\otimes n}\cong \C^{d^n}$. 
In quantum computation it is usual
to fix an orthonormal basis 
$\ket{0},\ldots,\ket{d-1}$ of
$\C^d$. An orthonormal basis of
$\C^{d^n}$ naturally corresponding
to this basis consists of vectors
of the form $\ket{i_1}\otimes\cdots\otimes\ket{i_n}$. 
This basis is called the computational basis.
The space $(\C^d)^{\otimes n}$ is
called an $n$-\qudit{} quantum system and
the the factors of the $n$-fold tensor
product $(\C^d)^{\otimes n}$ are referred
as the \qudits{} of the system. 

An $n$-\qudit{} quantum operation
(or gate) is a unitary transformation 
acting on the $n$-\qudit{} states,
i.e., an element of the unitary group $U_{d^n}$.
As in quantum computation, states which 
are scalar multiples of each other are 
considered equivalent, quantum operations
are also understood projectively. In particular,
for every $u\in U_{d^n}$, the normalized
operation $(\det u)^{-1/{d^n}}\cdot u$
represents the same gate as $u$
(here $\alpha^{1/d^n}$ stands for any $d^n$th
root of a complex number $\alpha$). 

Let $\gates\subset U_{d^n}$ be a (finite)
collection of $n$-\qudit{} quantum gates.
We say that $\gates$ is a {\it complete}
set of $n$-\qudit{} gates if a scalar multiple
of every $n$-\qudit{} operation from $U_{d^n}$, 
can be approximated with an arbitrary
precision by a product of operations from
$\gates$.
In other words, $\gates$ is
complete if the semigroup
of $U_{d^n}$ generated by $\gates$ 
and the unitary scalar matrices
is dense in $U_{d^n}$. 
The latter condition, because of
compactness, is equivalent to saying that
the {\it group} generated by $\gates$ 
and the unitary scalar matrices
is dense in $U_{d^n}$.

Note that in the quantum computation
literature complete sets of gates are frequently 
called universal. In this paper, partly following 
the terminology of \cite{Jeandel04},
we reserve the term {\it universal} for
expressing a weaker version discussed below. 

For $N\geq n$ we can view $(\C^d)^{\otimes N}$ as a
bipartite system 
$(\C^d)^{\otimes n}\otimes (\C^d)^{\otimes N-n}$
and let an $n$-\qudit{} gate $u$ act on
the first part only. Formally,
the $N$-\qudit{} extension $u_N$ of $u$ 
is the operation $u\otimes I$ where
$I$ stands for the identity of $(\C^d)^{\otimes N-n}$.
For an $n$-\qudit{} gate set $\gates$ the gate set $\gates_N$
is the collection of the extensions
of gates from $\gates$ obtained this way:
$\gates_N=\{u_N|u\in\gates\}$.

More generally, we can extend an $n$-\qudit{}
gate $u$ to $N$ \qudits{} by selecting an embedding
$\mu$ of $\{1,\ldots,n\}$ into $\{1,\ldots,N\}$
and let act $u$ on the components indexed by 
$\mu(1),\ldots,\mu(n)$ (in this order) and leave
the rest "unchanged". It will be convenient to
formalize this in terms of permutations of 
the \qudits{} of the larger system as follows. 
Each permutation from the symmetric group $S_N$ 
acts on $(\C^d)^{\otimes N}$ by permuting
the tensor components. For an $N$-\qudit{} 
gate $v$ and $\sigma\in S_N$
the operation $v^\sigma=\sigma v\sigma^{-1}$ 
is also a quantum gate which can be considered
as the gate $v$ with "fans" permuted by $\sigma$.
We denote by $\gates^N$ the collection of gates obtained
from gates in $\gates_N$ this way:
$\gates^N=\{u_N^\sigma|u\in\gates, \sigma\in S_N\}$.

We say that for $N\geq n$ the $n$-\qudit{}
gate set $\gates$ is {\it $N$-universal} if 
$\gates^N$ is complete. The collection
$\gates$ is called $\infty$-universal 
or just universal, for short, if there
exists $N_0\geq n$ such that $\gates$
is $N$-universal for every $N\geq N_0$.
It turns out that for $n\geq 2$, every complete
$n$-\qudit{} gate is $N$-universal for every $N\geq n$. 
This claim follows from the fact that the Lie algebra $su_{d^N}$ 
is generated by $su_{d^2}^N=
\{(u\otimes I)^\sigma|u\in su_{d^2},\sigma\in S_N\}$. 
This 
is shown in \cite{DiVi95} for $d=2$ but essentially
the same proof works for $d>2$ as well.

Hence an $n$-\qudit{} gate set $\gates$ is 
universal if and only if there exists an 
integer $N\geq n$ such that $\gates$ is $N$-universal.
On the other hand, no $1$-\qudit{} gate set can be
universal as the resulting group preserves
the natural tensor decomposition. 

\medskip

Completeness of a gate set can be decided 
by computing the (real) Zariski closure of the group 
generated by the gates using the method in \cite{DJK05}. 
A polynomial time algorithm for gates defined over
a number field is given in \cite{Jeandel04, Jeandel-th}.
Reducing the problem of universality to completeness
requires a bound for the smallest $N$ such that
a universal set of gates is $N$-universal.
In \cite{Jeandel04, Jeandel-th} Jeandel
gives a $6$-qubit gate set which is $9$-universal 
but not $6$-universal and it is explained
how to extend this example to a gate set over
$2^k+2$ qubits which is $2^{k+1}+1$-universal
but not $2^{k+1}-2$-universal where $k$
is an integer greater than $1$. 
(A qubit is a \qudit{} with $d=2$.) Our main
result is the following.

\begin{theorem}
\label{univ-thm}
Let $\gates$ be an $n$-\qudit{}
gate set where $n,d\geq 2$. 
Then $\gates$ is universal 
if and only if it is $N$-universal 
for some integer $N\leq d^{8}(n-1)+1$.
\end{theorem}

\medskip

Our main technical tool, a criterion for completeness
based on invariants of groups, is
given in Section~\ref{completeness-section}.
It can be considered as a "more algebraic" variant of 
Jeandel's criterion given in \cite{Jeandel04, Jeandel-th}. 
Correctness 
is a consequence 
of a recent result of Guralnick and Tiep
stating that certain low degree invariants distinguish 
the special linear group from its closed (in particular finite)
 subgroups. Needless 
to say, the proof of the applied result heavily uses
the classification of finite simple groups and their 
representations. 

We prove Theorem~\ref{univ-thm}
in Section~\ref{universality-section}. The outline
of the proof is the following. We relate
polynomial ideals to gate sets. The completeness
criterion gives that the Hilbert polynomial
of the ideal corresponding to a universal gate set 
must be the constant polynomial $24$. Our result is then a
consequence of Lazard's bound on the regularity
of Hilbert functions of zero dimensional ideals.

\section{Completeness}
\label{completeness-section}

In Jeandel's work \cite{Jeandel04,Jeandel-th}, 
testing gate sets for completeness
is based on the following observation.

\begin{fact}
\label{alternative-fact}
Let $d\geq 2$ and let $G$ be subgroup of $SU_{d^N}$. 
Assume further the real vector
space $su_{d^N}$ (the Lie algebra of $SU_{d^N}$)
consisting of the traceless skew
Hermitian $d^N\times d^N$ matrices is an irreducible
$\R G$-module under the conjugation action by elements of $G$.
Then $G$ is either finite or dense in $SU_{d^N}$.
\end{fact}

Therefore if $\gates$ is a finite collection
of normalized gates then testing $\gates$ for completeness
amounts to testing irreducibility of $su_{d^N}$ 
under conjugation of elements of $\gates$
and to testing if the linear group generated by
$\gates$ is finite. Informally, we are going to replace
the latter test with a test similar to the first one.

Set $V=\C^{d^N}$, the complex column vectors of
length $d^N$. The vector space $V$ is a left $\C G$-module
for every linear group $G\leq GL_{d^N}(\C)$. The dual
space $V^*=\Hom_\C(V,\C)$ is a right $\C G$-module. It can be made
a left $\C G$ module by letting $u^{-1}$ act in place
of $u$. This module (denoted also by $V^*$) is
called the module contragradient to $V$. In terms of
matrices, the contragradient matrix representation can be
obtained by taking the inverse of the transpose of
the original matrix representation. Note that for 
$u\in U_{d^N}$ the matrix of $u$ in the contragradient
representation will be simply the complex conjugate 
of the matrix of $u$. 

For every positive integer $k$ and $G\leq GL_{d^N}(\C)$
we define the quantity $\Monodr_{2k}(G)$ as
$$\Monodr_{2k}(G)=\dim_\C\Hom_{\C G}((V\otimes V^*)^{\otimes k},\C).$$
Recall that for a left $\C G$-module $W$
$$\Hom_{\C G}(W,\C)=\{f\in W^*|f(gw)=f(w)\mbox{~for every~}g\in G,w\in W\}.$$
Note that if a finite set $\gates$ generates a dense subgroup of $G$
and $B$ is a basis of $W$ then 
\begin{equation}
\label{hom-eq}
\Hom_{\C G}(W,\C)=
\{f\in W^*|f(gw)=f(w)\mbox{~for every~}g\in \gates,w\in B\},
\end{equation}
and hence (a basis of) the space $\Hom_G(W,\C)$ can
be computed by solving a system of linear equations.

Also note that $V\otimes V^*\cong \End_\C(V)$
and $\Monodr_2(G)$ is the dimension
of the centralizer of $G$ (in $\End_\C(V)$). 
In particular, $\Monodr_2(G)=1$ if and only
if $V$ is an irreducible $\C G$-module.
Similarly, $\Monodr_4(G)$ is the dimension of the
centralizer of the conjugation action of $G$
on $d^N\times d^N$ complex matrices. 

M.~Larsen observed that if ${\cal G}$ is 
the entire complex linear group $GL_{d^N}(\C)$, or the
complex orthogonal group or the complex symplectic group
and $G$ is a Zariski closed subgroup of ${\cal G}$
such that the connected component of the identity
in $G$ is reductive (including the case when this component is
trivial) and $\Monodr_4(G)=\Monodr_4({\cal G})$ then 
either $G$ is finite or $G\geq [{\cal G},{\cal G}]$.
(Notice that Fact~\ref{alternative-fact} can be viewed
as the unitary analogue of Larsen's alternative.)
Larsen also conjectured that for a finite subgroup
$G< {\cal G}$ we have $\Monodr_{2k}(G)>\Monodr_{2k}({\cal G})$
with some $k\leq 4$. Recently R.~M.~Guralnick and P.~H.~Tiep 
\cite{GT05},
using the classification of finite simple groups and
their irreducible representations, settled
Larsen's conjecture. The conjecture holds basically true,
there are only two exceptions. In any case, 
$\Monodr_{2k}(G)>\Monodr_{2k}({\cal G})$ with some $k\leq 6$.
The following statement is an easy consequence
of the results from \cite{GT05}. In order to
shorten notation, for a collection 
$\gates\subseteq U_{d^N}$ we define
$\Monodr_{2k}(\gates)$ as $\Monodr_{2k}(G)$ where
$G$ is the smallest closed subgroup
of $U_{d^N}$ containing $\gates$ (in the
norm topology). Also, in view (\ref{hom-eq})
and the comment following it, computing
$\Monodr_{2k}(\gates)$ can be accomplished
by computing the rank of 
a $d^{N^{2k}}$ by $|\gates|d^{N^{2k}}$
matrix
if $\gates$ is finite.

\begin{proposition}
\label{completeness-prop}
Assume that $d^N>2$ and let $\gates\subset U_{d^N}$.
Then $\gates$ is complete if and only if
$\Monodr_8(\gates)=\Monodr_8(GL_{d^N}(\C))$. If $d^N=2$ then
the necessary and sufficient condition for completeness
is $\Monodr_{12}(\gates)=\Monodr_{12}(GL_{d^N}(\C))$.
\end{proposition}

\begin{proof}{Proof}
We only prove the first statement,
the second assertion can be verified
with a slight modification of the arguments.
Let $G$ be the smallest closed subgroup
of $U_{d^N}$ containing $\gates$ (in the norm
topology). We replace 
each $u\in G$ with its normalized version 
$\det^{-1}u\cdot u$. In this way we achieve
that $G$ is a closed subgroup of $SU_{d^N}$.
As the action of $\det^{-1}u\cdot u$
is the same as that of $u$ on
$V^{\otimes k}\otimes {V^*}^{\otimes k}$,
this change does not affect the quantities
$\Monodr_{2k}(G)$. If $\gates$ is complete
then $G=SU_{d^N}$. Therefore the Zariski closure
of $G$ in $GL_{d^N}(\C)$ (over the complex numbers)
is $SL_{d^N}(\C)$ and
hence $\Monodr_{2k}(G)=\Monodr_{2k}(SL_{d^N}(\C))=
\Monodr_{2k}(GL_{d^N}(\C))$ for every $k$. 
This shows the "only if" part.

To prove the reverse implication,
assume that $\Monodr_8(G)=\Monodr_8(GL_{d^N}(\C))$.
By Lemma~3.1 of \cite{GT05}, 
$\Monodr_{2k}(G)=\Monodr_{2k}(GL_{d^N}(\C))$
for $k=1,2,3$ as well. In particular,
$\Monodr_4(G)=\Monodr_4(GL_{d^N}(\C))=2$. Notice that 
$G$ is a compact Lie group therefore
every finite dimensional representation of
$G$ is completely reducible. Hence the
the conjugation action of $G$ on 
$d^N\times d^N$ matrices has two
irreducible components: one consists
of the scalar matrices the other one is
the Lie algebra $sl_{d^N}(\C)$ of
traceless matrices. As a real
vector space, $sl_{d^N}(\C)$
is the direct sum of $su_{d^N}$
and $i\cdot su_{d^N}$ (here $i=\sqrt{-1}$). Both subspaces
are invariant under the action
of $U_{d^N}$, therefore they
are $\R G$-submodules and multiplication
by $i$ gives an $\R G$-module isomorphism between
them. It follows that $su_{d^N}$ must
be an irreducible $\R G$-module. Hence
by Fact~\ref{alternative-fact}, either
$G=SU_{d^N}$ or $G$ is finite. In the first
case $\gates$ is complete. In the second case
we can apply the results of \cite{GT05}. By
Theorems~1.4 and~2.12 therein, $G$ must be
$SL_2(5)$ and $d^N=2$. This contradicts
the assumption $d^N>2$.
\end{proof}

\section{Universality}
\label{universality-section}

We begin with a lemma which establishes
a condition for $N$-universality
which suits better our purposes
than the original definition.

\begin{lemma}
\label{univ-generators-lemma}
Let $d>1$ and $\gates$ be an $n$-\qudit{} gate set, let $N\geq n$
and let $\Sigma$ be an arbitrary generating set for $S_N$.
Then $\gates$ is $N$-universal if and only if
$\gates_N\cup \Sigma$ is complete.
\end{lemma}

\begin{proof}{Proof}
Let $H$ resp.~$G$ denote the closure of the subgroup
of $SU_{d^N}$
generated by the normalized gates from $\gates^N$ and 
$\gates_N\cup \Sigma$, respectively.
As $\gates^N$ is in the subgroup generated by
$\gates\cup \Sigma$, the group $H$ is a subgroup
of $G$ and hence the "only if" part of the statement
is obvious. 
On the other hand, $H$ is closed
under conjugation by the elements of $\gates\cup S_N$,
therefore $H$ is a closed normal subgroup of $G$.
Assume that $\gates_N\cup \Sigma$ is complete, i.e.,
$G=U_{d^N}$. Then $\gates_N$ must 
contain at least one non-scalar matrix since 
otherwise $G$ would be finite
(every matrix in $G$ would be a permutation, multiplied
by a $d^N$th root of unity). 
Therefore $H$ is a normal subgroup of $SU_{d^N}$ 
containing a non-scalar matrix. Because
of simplicity of $PSU_{d^N}$ this implies
$H=SU_{d^N}$, that is, $\gates^N$ is complete.
\end{proof}

By Lemma~\ref{univ-generators-lemma}, we
can consider gate sets on $N$ \qudits{}
which consist of two parts. The gates
in the first part act on the first
$n$ \qudits{} while the rest consists
of permutations. We exploit this property in 
Subsection~\ref{ideal-ss}, where
we relate polynomial ideals to
such a sequence of gate sets where
$N$ varies. We finish the proof
of Theorem~\ref{univ-thm} in 
Subsection~\ref{proof-ss} by observing
that the sequence $\Monodr_8$ for
letting an $n$-\qudit{} gate set together with
the symmetric group $S_N$ act
on $(C^{d})^{\otimes N}$ ($N=n,n+1,\ldots$)
take the same values as the Hilbert function
of the corresponding ideal.

\subsection{The ideal of a gate set}

\label{ideal-ss}

In this subsection $W=\C^m$ for
some integer $m>0$ and $G$ is a
subgroup of $GL(W^{\otimes n})$.
For every $N\geq n$ we 
establish a relation between 
$\Hom_{\langle G,S_n\rangle}(W^{\otimes n},\C)$ 
and $\Hom_{\langle G\otimes I,S_N\rangle}(W^{\otimes N},\C)$. Here
$S_N$ denotes the subgroup of $GL(W^{\otimes N})$
consisting of the permutations of tensor components
and $I$ stands for the identity on $W^{\otimes (N-n)}$. 

We work with the tensor algebra
$T=\oplus_{j=0}^\infty W^{\otimes j}$
of $W$. We use some elementary properties
of $T$ and its substructures. Most of the proofs
can be found in Section~9 of \cite{Greub}.
We say that an element $w$ of $T$ is
homogeneous of degree $j$ if $w\in W^{\otimes j}$.
If we fix a basis
$w_1,\ldots,w_m$ of $W$, then a basis of $T$
consists of the non-commutative monomials
of the form $w_{i_1}\otimes \cdots \otimes w_{i_j}$ and
$T$ can be interpreted as the ring of non-commutative polynomials
in $w_1,\ldots,w_m$ over $\C$. In this interpretation,
for every $j\geq 0$ the elements
of $W^{\otimes j}$ are identified with the homogeneous 
non-commutative polynomials of degree $j$. 
A right (or two sided) ideal $J$ of $T$ is 
called graded if $J$ equals the sum $\oplus_{j=0}^\infty J^j$ 
where $J^j=W^{\otimes j}\cap J$. The component $J^j$ is called 
the degree $j$ part of $J$. It turns out that a right (resp.~two-sided)
ideal $J$ of $T$ is graded if and only if there is a
set of homogeneous elements of $J$ which generate $J$ 
as a right (resp. two-sided) ideal.

Let $M$ be the two-sided ideal of $T$ generated by 
$w_i\otimes w_j-w_j\otimes w_i$ ($i,j\in\{1,\ldots m\}$),
and let $\phi:T\rightarrow R=T/M$ be the natural map. 
Then $M$ is a graded ideal with degree $j$ parts
$M^j$ which are spanned by $w_{i_1}\otimes \cdots \otimes w_{i_j}-
w_{i_{\sigma(1)}}\otimes \cdots \otimes w_{i_{\sigma(j)}}$ where
$(i_1,\ldots i_j)\in\{1\ldots,m\}^j$ and $\sigma\in S_j$.
The factor algebra $R$ is called the symmetric algebra of $W$.
Set $x_i=\phi(w_i)$ for $i=1,\ldots,m$. Then $R$ is identified with
the (commutative) polynomial ring $\C[x_1,\ldots,x_m]$.
The image of $R^j$ of $W^{\otimes j}$ under $\phi$
is the $j$th symmetric power of $W$. In interpretation
of $R$ as polynomial ring, $R^j$ consists 
of the homogeneous polynomials of degree $j$.

For a subspace $L$ of $(W^{\otimes N})^*$
we denote by $L^\perp$ the subspace of
$W^{\otimes N}$ annihilated by $L$:
$L^\perp=\{w\in W^{\otimes N}|l(w)=0\mbox{~for every~}l\in L\}$.
Because of duality, $\dim L=\dim (W^{\otimes N}/L^\perp)$
and $(L_1\cap L_2)^\perp=L_1^\perp+L_2^\perp$.
In particular,
$\Hom_{\langle G\otimes I\cup S_N\rangle}
(W^{\otimes N},\C)^\perp=
\Hom_{G\otimes I}(W^{\otimes N},\C)^\perp+
\Hom_{S_N}(W^{\otimes N},\C)^\perp$.

As $\Hom_{G\otimes I}(W^{\otimes N},\C)
=\Hom_G(W^{\otimes n},\C)\otimes (W^{\otimes(N-n)})^*$, 
we obtain that
$\Hom_{G\otimes I}(W^{\otimes N},\C)^\perp
=\Hom_G(W^{\otimes n},\C)^\perp \otimes W^{\otimes(N-n)}$, 
in other words, the space
$\Hom_{G\otimes I}(W^{\otimes N},\C)^\perp$
is the degree $N$ part of
the right ideal $H(G)$ in $T$ generated by 
$\Hom_{G}(W^{\otimes n},\C)^\perp$.

The space $\Hom_{S_N}(W^{\otimes N},\C)$
corresponds the symmetric $N$-linear functions,
i.e., it consists of the linear functions $W^{\otimes N}\rightarrow \C$ 
which take identical values on
$w_{i_1}\otimes \cdots \otimes w_{i_N}$ and
$w_{i_{\sigma(1)}}\otimes \cdots \otimes w_{i_{\sigma(N)}}$
for every permutation $\sigma \in S_N$. Therefore
$\Hom_{S_N}(W^{\otimes N},\C)^\perp$ coincides
with the degree $N$ part $M^N$
of the ideal $M$.

We obtain that
$\Hom_{\langle G\otimes I\cup S_N\rangle}
(W^{\otimes N},\C)^\perp$ is the degree
$N$ part of $H(G)+M$. As
$H(G)$ is a right ideal and $M$ is an ideal
in $T$ with $R=T/M$ commutative, $H(G)+M$
is an ideal in $T$ containing $M$. Setting
$J(G)=\phi(H(G)+M)$ we conclude
that for every $N\geq n$, 
$J^N(G)=\phi(\Hom_{\langle G\otimes I\cup S_N\rangle}
(W^{\otimes N},\C)^\perp)$ is
the degree $N$ part
of $J(G)$. Furthermore, $J(G)$ is the ideal
of the commutative polynomial ring $R$
generated by $J^n(G)$ and 
$$\dim \Hom_{\langle G\otimes I\cup S_N\rangle}
(W^{\otimes N},\C)=\dim(R^N/J^N(G)).$$

\subsection{The proof of Theorem~\ref{univ-thm}}

\label{proof-ss}

Let $n,d\geq 2$, let $\gates\subseteq GL((\C^d)^{\otimes n})$
and let $G$ be the subgroup of $GL(\C^d)$
generated by $\gates$. For every 
integer $N\geq n$, we consider the
$G$-module $V=(\C^d)^{\otimes N}$
where the action of $G$ is given
by $G\otimes I$ (here $I$ is the identity
on $V^{\otimes (N-n)}$). We set
$W=(\C^d)^{\otimes 4}\otimes ((\C^d)^*)^{\otimes 4}$
and, with some abuse of notation, consider
$G$ as a subgroup of $GL(W^{\otimes n})$.
For every $N\geq n$ we have the $G$-module isomorphism
$V^{\otimes 4}\otimes (V^*)^{\otimes 4}\cong W^{\otimes N}$
where the action of $G$ on the right hand side
is $G\otimes I$ (this time $I$ is the identity
on $W^{\otimes (N-n)}$). Applying the notation
and observations of the preceding subsection in this context,
we obtain that 
$$\Monodr_8(\langle G\otimes I\cup S_N\rangle)=
\dim (R^N/J^N(G))$$
for every $N\geq n$.

First we consider the full linear group
$GL_{d^n}(\C)$. The $n$-universality of 
$U_{d^n}$ for $n\geq 2$ gives
$\dim (R^N/J^N(GL_{d^n}(\C))=\Monodr_8(GL_{d^N}(\C))$.
From invariant theory it is known that 
$\Monodr_8(GL_{d^N}(\C))=4!=24$, see \cite{Weyl46}.

Now consider an arbitrary gate set $\gates\subseteq U_{d^n}$
and let $G\leq GL_{d^n}(\C)$ the group generated by 
$\gates$. The preceding discussion and
Proposition~\ref{completeness-prop}
give that $\gates$ is universal
if and only if $\dim (R^N/J^N(G))=24$ for sufficiently large
degree $N$.

The ideal $J(G)$ is an ideal of $R=\C[x_1,\ldots,x_m]$ 
generated by homogeneous polynomials of degree $n$.
In the context of polynomial rings,
graded ideals are called homogeneous.
That is, an ideal $J$ of the polynomial ring $R$ 
is called homogeneous
if $J$ is the direct sum its homogeneous components
$J^j=R^j\cap J$; and an ideal generated by homogeneous
polynomials is homogeneous. The {\it Hilbert function}
of the homogeneous ideal $J$ is given as
$j\mapsto h_J(j)=\dim R^j/J^j$. It turns out that
the Hilbert function is ultimately a polynomial:
there is a polynomial $p_J$ (in one variable)
and an integer $N$ such that $h_J(j)=p_J(j)$
for $j\geq N$. The smallest $N$ with this 
property is called the regularity of the
Hilbert function of $J$. The degree of the
Hilbert polynomial is the {\it dimension}
of $J$. (Actually, it is the dimension of
the projective variety consisting of the
common projective roots of the polynomials
in $J$.) 

The discussion above shows that 
the Hilbert polynomial of the ideal
$J(G)$ corresponding to a universal
gate set is the constant 24. 
In particular, the dimension of $J(G)$
is zero. In \cite{Lazard81}, D.~Lazard
proved that the regularity of
the Hilbert function of a zero dimensional ideal
in $\C[x_1,\ldots,x_m]$ generated by 
homogeneous polynomials of degree
$n$ is bounded by $mn-m+1$. 
From this, the proof of Theorem~\ref{univ-thm}
is finished by observing that
the smallest $N$ for which $\gates$ is
$N$-universal coincides with the regularity
of the Hilbert function of $J(G)$.

\section{Concluding remarks}

Very probably the bound proved in Theorem~\ref{univ-thm}
is not tight. However, for fixed $d$ it is linear in
$n$ and Jeandel's construction discussed
in the introduction shows that in fact the smallest
$N$ such that a universal $n$-qubit gate set is $N$-universal
can be at least $2n-6$. 

Proving better upper
bounds would require deeper knowledge
of subspaces of ${V^*}^{\otimes 4}\otimes V^{\otimes 4}$
which occur as $\Hom_G(V^{\otimes 4}\otimes {V^*}^{\otimes 4},\C)$
for $G\leq GL(V)$. Using the isomorphism 
$\Hom_G(V^{\otimes 4}\otimes {V^*}^{\otimes 4},\C)
\cong\End_G(V^{\otimes 4})$, a natural restriction is
that these subspaces must be subalgebras of 
$\End_{\C}(V^{\otimes 4})$. However, it is not obvious
how to exploit this fact.

\medskip

Effectiveness and complexity of algorithms for
testing completeness and universality based on
Proposition~\ref{completeness-prop}, Theorem~\ref{univ-thm}
and Lemma~\ref{univ-generators-lemma}
depend on the computational model and on the way
how the input gate set is represented. In the Blum--Shub--Smale
model, if the input gates are given as arrays of $n\times n$
complex numbers, the completeness test can be accomplished in
polynomial time. With the same assumption on the input, for
constant $d$ (e.g., for qubits or qutrits) even universality 
can be tested in polynomial time. Similar results can
be stated for Boolean complexity if the entries of the 
matrices representing the input gates are from an algebraic 
number field. Even it is decidable if there is a non-universal
gate set which is $\epsilon$-close to a given collection
of gates in the Hadamard norm of matrices. Indeed, 
existence is equivalent to solvability of a (huge) 
system of polynomial equations and inequalities
over the real numbers. Of course, this straightforward
method is far from practical.

\paragraph*{Acknowledgments}
The author is grateful to Emmanuel Jeandel, Lajos R\'onyai
and Csaba Schneider and to an anonymous referee for their
 useful remarks and suggestions.

\end{document}